\documentclass[a4]{nature}

\usepackage{amsmath}
\usepackage{graphicx,color}
\newcommand{\mNa}{m}
\newcommand{\drag}{{\it constant force}}
\newcommand{\decay}{{\it oscillation}}

\title{Spin Drag in a Bose Gas}
\author{S.B. Koller$^{1}$, A. Groot$^{1}$, P.C. Bons$^{1}$, R.A. Duine$^{2}$, H.T.C. Stoof$^{2}$\& P. van der Straten$^{1}$}
\begin{document}
\maketitle
\begin{affiliations}
\item Nanophotonics, Debye Institute, Utrecht University, PO Box 80,000, 3508 TA Utrecht, The Netherlands
\item Institute for Theoretical Physics, Utrecht University, PO Box 80,000, 3508 TA Utrecht, The Netherlands
\end{affiliations}

\begin{abstract} 
It is well known that the charge current in a conductor is proportional to the applied electric field. This famous relation, known as Ohm's law, is the result of relaxation of the current due to charge carriers undergoing collisions, predominantly with impurities and lattice vibrations in the material. The field of spintronics\cite{Wolf,Mae2002,Zutic,Aws2007,Ohn2010}, where the spin of the electron is manipulated rather than its charge, has recently also led to interest in spin currents. Contrary to charge currents, these spin currents can be subject to strong relaxation due to collisions between different spin species, a phenomenon known as spin drag\cite{PhysRevB.68.045307}. This effect has been observed for electrons in semi-conductors\cite{Weber} and for cold fermionic atoms\cite{DeMarco,Sommer}, where in both cases it is reduced at low temperatures due to the fermionic nature of the particles.  Here,  we perform a transport experiment using ultra-cold bosonic atoms and observe spin drag for bosons for the first time. By lowering the temperature we find that spin drag for bosons is enhanced in the quantum regime due to Bose stimulation, which is in agreement with recent theoretical predictions\cite{Driel}. Our work on bosonic transport shows that this field may be as rich as transport in solid-state physics and may lead to the development of advanced devices in atomtronics\cite{Holland}.
\end{abstract}

\section{Body}

\begin{figure}
\begin{center}
\def\svgwidth{\textwidth}
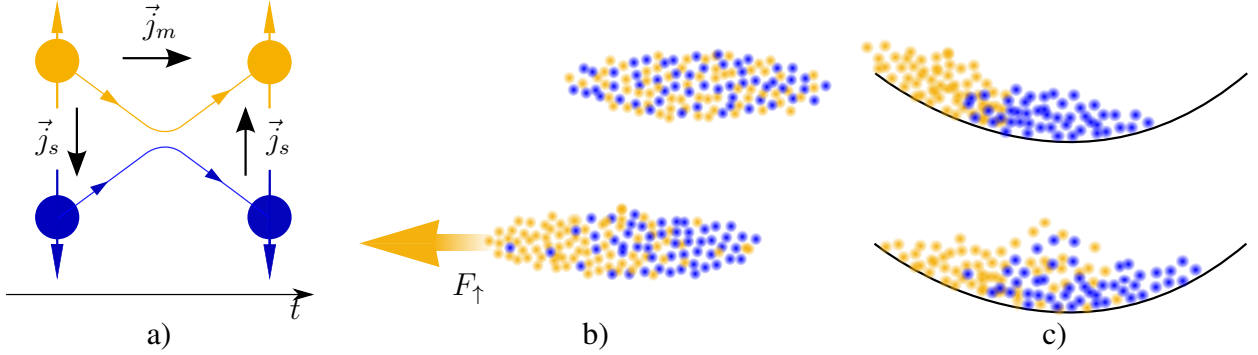
\end{center}
\caption{{\bf Schematical representation of the experimental methods} -- a) Collision of two atoms with different spins\cite{Weber}. The mass current $\vec{j}_m$ in this process as shown by the horizontal arrow is conserved, whereas the spin current $\vec{j}_s$ as shown by the vertical arrow is not conserved. In this example the spin current is reversed. b) Schematic representation of the measurement: Two spin species ({\it blue}, {\it yellow}) are prepared and a force acts on only one spin species ({\it yellow}). Due to the spin drag the other spin species ({\it blue}) is dragged along.    \label{fg:schematics}}
\end{figure}

% Introduction
Ohm's law can be written as $\vec{j}=\vec{E}/\rho$, showing the simple linear relation between the applied electric field $\vec{E}$ and the current density $\vec{j}$. The electrical resistivity $\rho$ comprises a wealth of information on the material and its careful measurement over the last century in solid-state physics has introduced new intriguing phenomena, like high-temperature superconductivity\cite{Bednorz} and the integer\cite{Klitzing} and fractional\cite{Tsui} quantum Hall effect. For the case of spin currents in two-component cold-atom systems with pseudo-spin {\em up} and {\em down} a similar relation applies\cite{Driel}:
\begin{equation}
  \vec{j}_s = \frac{\vec{F}_\uparrow}{\rho_s}, \label{eq:equdrag}
\end{equation}
where the applied force $\vec{F}_\uparrow$ on particles with spin {\em up}, mass $m$ and density $n$ leads to a drift velocity difference $\Delta \vec{v}$ between the two spin species. This results in a spin-current density $\vec{j}_s=n \Delta \vec{v}$, which is inversely proportional to the  spin resistivity $\rho_s$. Because of the absence of impurities and an ionic lattice the spin resistivity $\rho_s=m\gamma/n$ is for ultra-cold atoms completely determined by the spin-drag rate $\gamma$. This rate is a measure for the transfer of momentum via interactions between the two spin species. The similarity between Ohm's law and Eq.~(\ref{eq:equdrag}) can be understood from the analogy between their underlying physical mechanism. In both cases the external force or field leads to an acceleration, which on average is interrupted after a time $\tau$ and this leads to a finite current. For  spin currents in ultra-cold atoms the particles with one spin collide with particles of another spin and these collisions do not conserve the spin current, as shown in  Fig.~\ref{fg:schematics}a. This is the microscopic origin of  spin drag. It is important to emphasize that in electronic systems spin-drag effects also exist, but are usually small compared to the effects of impurities and phonons~\cite{Weber,PhysRevB.68.045307}, whereas for ultra-cold atoms they are the only effect and therefore dominant. 
 
% How implemented in ultra-cold gases
To observe spin drag in trapped ultra-cold atomic quantum gases, we set atoms with pseudo-spin {\em up} into motion with respect to atoms with pseudo-spin {\em down} and observe, how atoms with spin {\em down} are dragged along with the atoms with spin {\em up}. In our first method, the \drag\ method, we apply a constant force $\vec{F}_\uparrow$ on only the spin species {\em up} as shown in Fig.~\ref{fg:schematics}b. After a short period a constant drift velocity difference $\Delta \vec{v}$ has developed between the two spin species, where the ratio of the applied force $\vec{F}_\uparrow$ and $\Delta \vec{v}$ is proportional to the drag rate $\gamma$. This method is well suited for a large drag rates, but for weak drag a steady state may not be achieved before the two clouds are spatially separated. Therefore for small drag rates we apply our second method, the \decay\ method. In this method the centers of the two spin species are first separated from each other and then  start to oscillate in the trap. The decay time of the relative oscillation is proportional to the drag rate. For both methods we have to detect the change in relative position of the clouds as a function of time. We apply a Stern-Gerlach technique\cite{Stenger} in a direction perpendicular to the applied force to spatially separate the spin species and image the clouds with absorption imaging.

% Experimental parameters
In the experiment we load up to $4.6 \times 10^8$ $^{23}$Na atoms into a cigar-shaped optical far-off-resonant trap (FORT) with trap frequencies of $\omega_{\mathrm{rad}}/2\pi$ = 835 Hz in the radial and $\omega _{\mathrm{ax}}/2\pi$ = 3.5 Hz in the axial direction. The trap is tight in the radial direction to obtain a large density, thereby producing a collisionally opaque and hydrodynamic sample in the axial direction. The temperature is between 2--8 $\mu$K and is always kept above the critical temperature for Bose-Einstein condensation. The atoms in the trap are initially all in one spin state. For stability and to maximize the interspecies collisions we then flip about 50\% of the atoms to another spin state, creating an equal mixture of spin species  {\em up} and {\em down} (See Methods).

\begin{figure}
\begin{center}
\def\svgwidth{\textwidth}
{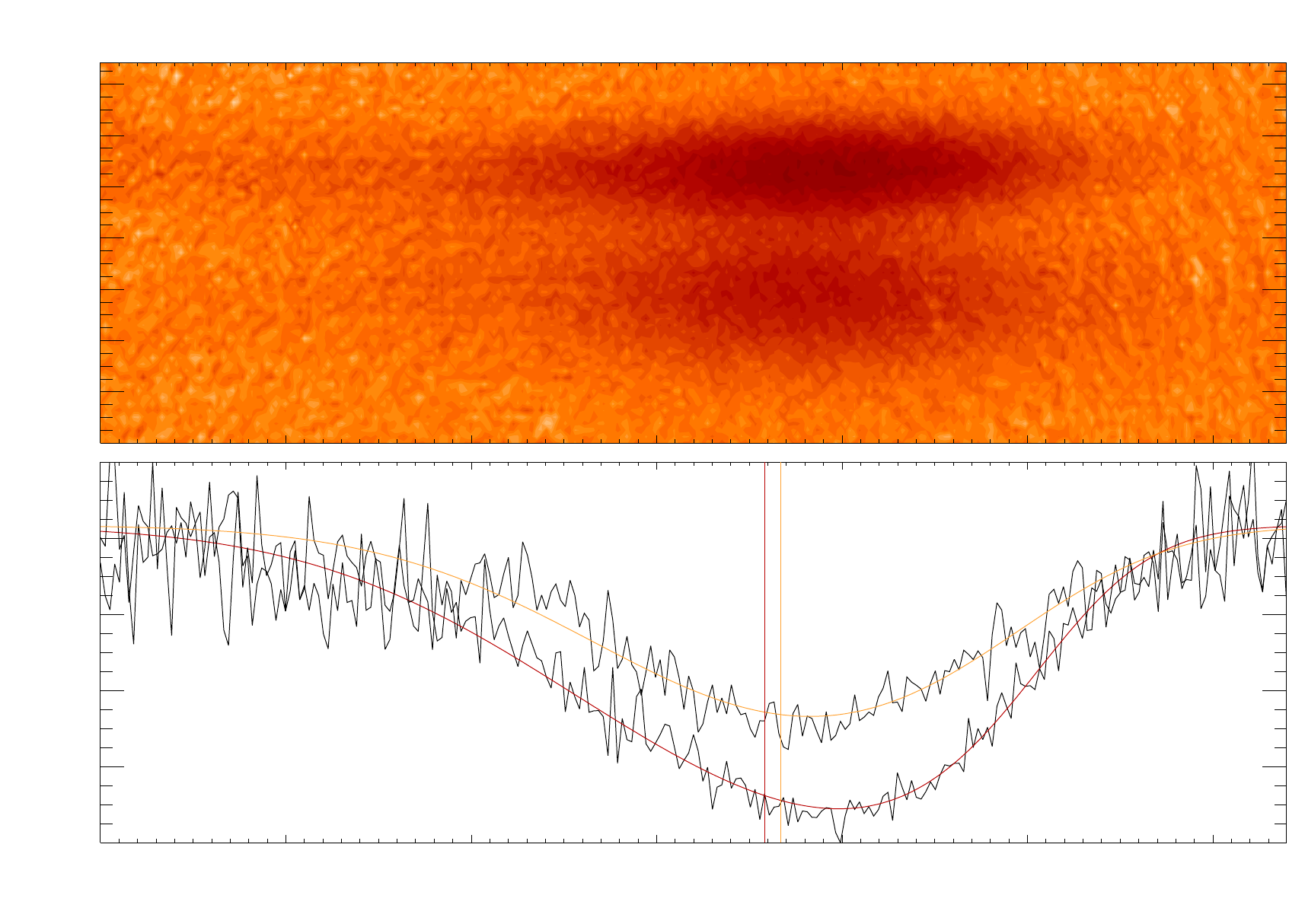}
\end{center}
\caption{{\bf Measurement for the \drag\ method} -- A single measurement for the \drag\ method on a cloud with $1.5\times10^8$ atoms at 4 $\mu$K. The spin species {\em up}  is accelerated with   1.01 m/s$^2$ to the left. a) Pseudo color image of the measured absorption along a line-of-sight through the two clouds. The upper cloud is in the spin {\em up} state, whereas the lower cloud is in the spin {\em down} state. The separation between the clouds results from the Stern-Gerlach technique {\em after} the spin drag to facilitate the imaging of the individual clouds. The difference in width of the clouds in the vertical direction develops during the detection stage.  c) Absorption of the two clouds in the horizontal direction, where the lower curve is for the upper cloud (spin {\em up}) and the upper curve is for the lower cloud (spin {\em down}). The two curves are obtained by making a cut through the 2D-image of Fig. \ref{fg:snapshot}a at the height of the respective clouds. The red/orange curves are for spin state {\em up}/{\em down} and represent a fit with the skewed Gaussian distributions for the individual clouds. The skew in the two distributions is a direct result of the drag of the upper cloud exerted on the lower cloud. Notice that the separation between the two clouds is  small compared to the width of the clouds. \label{fg:snapshot}}
\end{figure}

% Example of the drag method
In Fig.~\ref{fg:snapshot} we show an example of the \drag\ method. The force is applied on the spin species {\em up} such that this cloud starts to move to the left and due to spin drag the spin species {\em down} is dragged along.  Careful observation shows that the upper cloud of spin species {\em up} is skewed: The left wing of the cloud is longer than the right wing. This is a first, direct indication of spin drag, which is  shown more clearly in Fig.~\ref{fg:snapshot}b in the profiles along the pulling direction through the center of each cloud. The clouds are fitted to skewed Gaussian distributions as shown in Fig.~\ref{fg:snapshot}b to determine the center of mass. As indicated with the two vertical lines in Fig.~\ref{fg:snapshot}b, the difference in position between the two clouds is much smaller than the width of the clouds such that local effects can be neglected.

\begin{figure}
\begin{center}
\includegraphics[width=\textwidth]{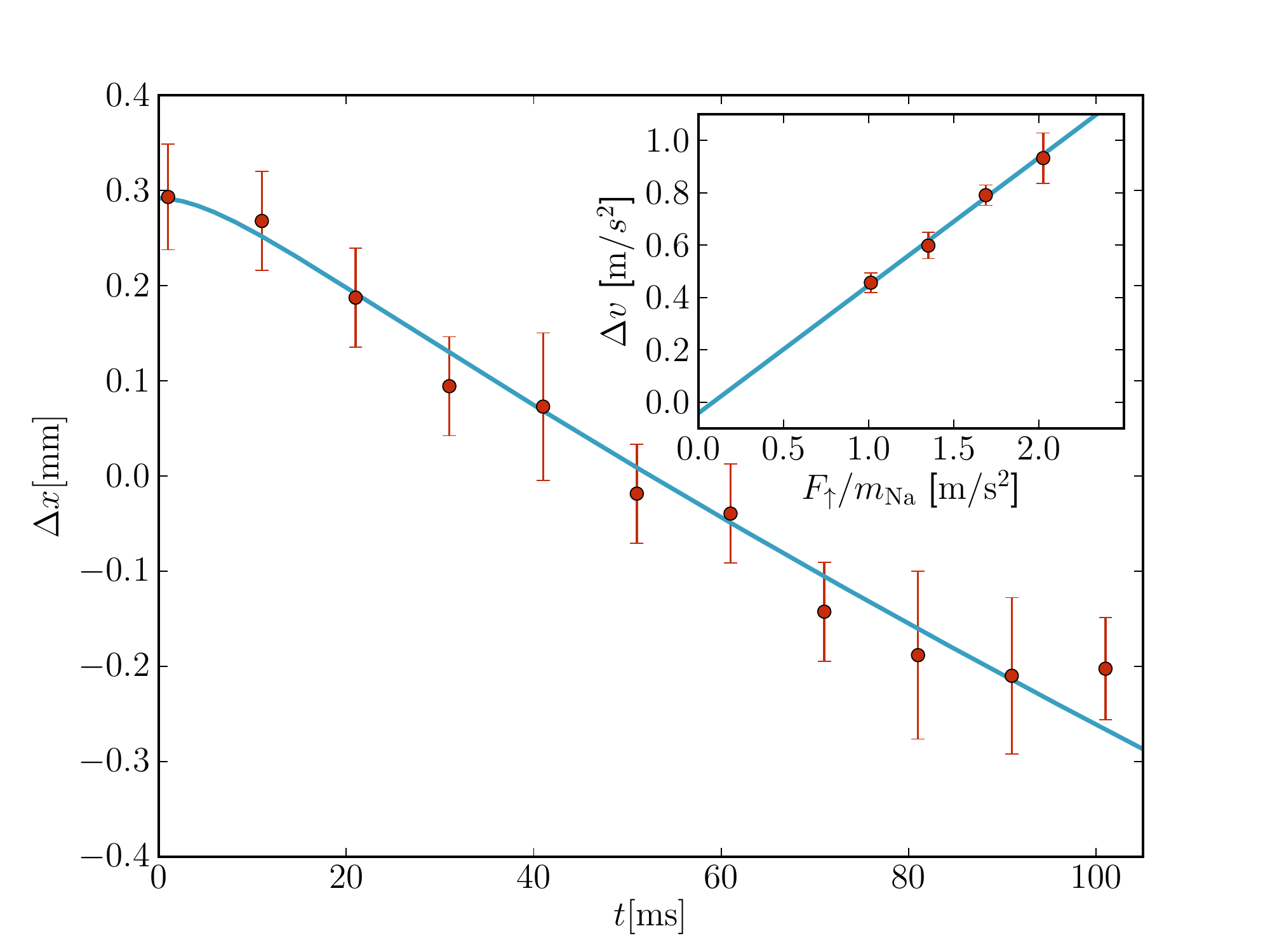}
\end{center}
\caption{{\bf Experimental results for the \drag\ method} -- Relative displacement $\Delta x$ between the position of the two clouds as a function of duration $t$ that the force is applied. The data points are indicated by the red dots and the blue line is a fit to the data using the solution of the coupled equations of motion of the two clouds. After a short period ($t <$ 5 ms) $\Delta x$ is linear in time, indicating that there is drag between the two clouds. All data points are offset due to an artefact in the Stern-Gerlach technique, which shifts after the spin drag the axial position of the two clouds with respect to each other by a constant amount. The inset shows the drift velocity  $\Delta v$ as a function of the strength of the force $F_\uparrow$. The straight line is a fit to the data, which shows that the experiments are in the linear-response regime for the drag, where Eq.~(\ref{eq:equdrag}) holds. \label{fg:fitcurve}}
\end{figure}

% Delta v vs. force (pulling)
To obtain the spin-drag rate this measurement is repeated for different magnitudes of the force and different durations during which the force is applied. In all cases we measure the relative displacement as shown in Fig.~\ref{fg:snapshot}b. The result for different duration $t$ is shown in Fig.~\ref{fg:fitcurve}. After a short time the difference in position between the two clouds is linear in time, indicating that there is a constant drift velocity $\Delta v$ between the two spin species. This is direct evidence for spin drag as the drag leads to friction between the two spin states. In the absence of drag the difference in position would show a quadratic dependence on time due to the ballistic motion of the atoms. The curvature for short times ($t <$ 5 ms) in Fig.~\ref{fg:fitcurve} is due to the fact that the two clouds are initially not in a steady state. However, the clouds are collisionally opaque, {\it i.e.}, they are in the hydrodynamical regime, and the two clouds reach  quickly a steady state. In our analysis we allow for this by solving the coupled equations of motion for the centers of the two clouds\cite{Duine} and the results are used to extract $\gamma$ from the data.

% Linear regime (pulling)
Equation~\ref{eq:equdrag} is only valid for small applied force ${F}_\uparrow$, such that the response of the clouds is still in the linear regime. Two of us\cite{Duine} showed that for ultra-cold atoms to be in the linear-response regime $\Delta v$ should be well below the thermal velocity $v_{\mathrm{th}}=\sqrt{{8 k_B T}/{\pi \mNa}}$, where $k_B$ is the Boltzmann constant and $T$ the temperature. In the experiments $v_{\mathrm{th}}$ is of the order of 6 cm/s, whereas typical values for $\Delta v$ are of the order of 1 mm/s. Thus we expect that our experiments are in the linear regime. In the inset of Fig.~\ref{fg:fitcurve} we have tested this assumption by measuring $\Delta v$ as a function of the acceleration $F_\uparrow/\mNa$. The line is a linear fit that shows good agreement with the data, which indicates that this measurement is indeed in the linear regime. Since we conducted this measurement with the lowest number of particles and thus lowest drag rate, this proves that all our measurements for the \drag\ method are in the linear-response regime.

\begin{figure}
\begin{center}
\includegraphics[width=\textwidth]{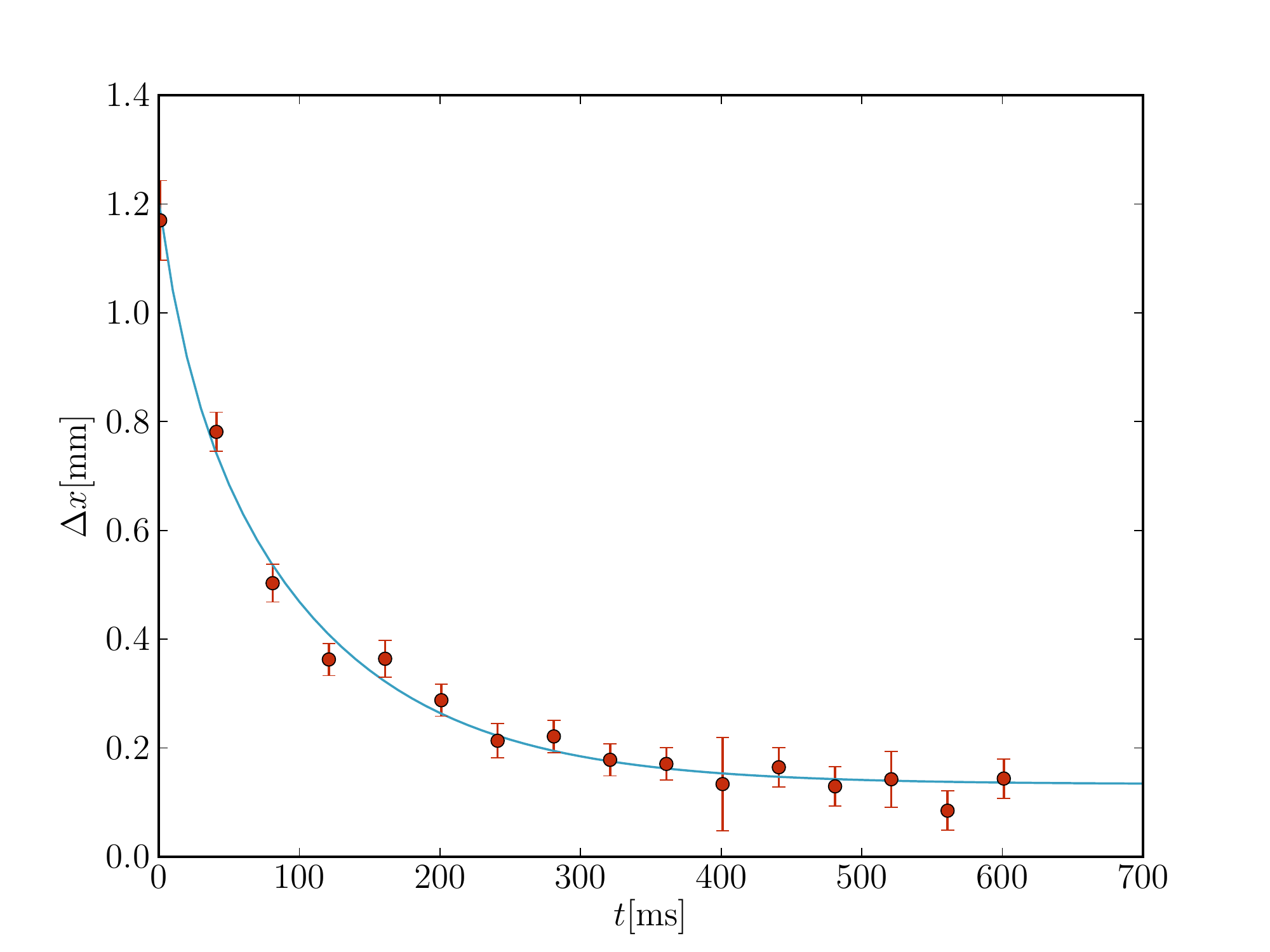}
\end{center}
\caption{{\bf Experimental results for the \decay\ method} -- The relative displacement $\Delta x$ in position between the two clouds for the \decay\ method as a function of time $t$ after the release of the atoms. The relative oscillation is overdamped, in which case the decay rate $\beta_d$ of the oscillation is given by $\beta_d = 2 \omega_{\rm ax}/\gamma$ with $\omega_{\rm ax}$ the oscillation frequency in the trap and $\gamma$ the drag rate. The red points are the data points and the blue curve is a fit to the data assuming exponential decay. All data points are offset due to an artefact in the Stern-Gerlach technique, which shifts after the oscillation the axial position of the two clouds with respect to each other by a constant amount. \label{fg:betavspull}}
\end{figure}

% decay method
As mentioned, for lower drag rates we applied the \decay\ method. Here, only 3--7$\times10^6$ atoms are loaded into the FORT. Atoms with spin {\em up} are spatially separated from atoms with spin {\em down} by exerting with a constant force on the spin species {\em up} for a period, after which the force is switched off. The atoms in spin state {\em up} start to oscillate in the trap and drag along the atoms in spin state {\em down}, which are initially at rest.   In Fig.~\ref{fg:betavspull} we show the relative displacement between the two clouds as a function of time after releasing the atoms. Since the drag rate $\gamma$ is much larger than the axial trap frequency $\omega_{\rm ax}$, the relative oscillation is overdamped and the decay rate is given by $\beta_d = 2 \omega_{\rm ax}{}^2/\gamma$, which allows us to extract the spin-drag rate $\gamma$ from the measurements. Even for the lowest drag rates the oscillation remains overdamped. 

\begin{figure}
\begin{center}
\includegraphics[width=\textwidth]{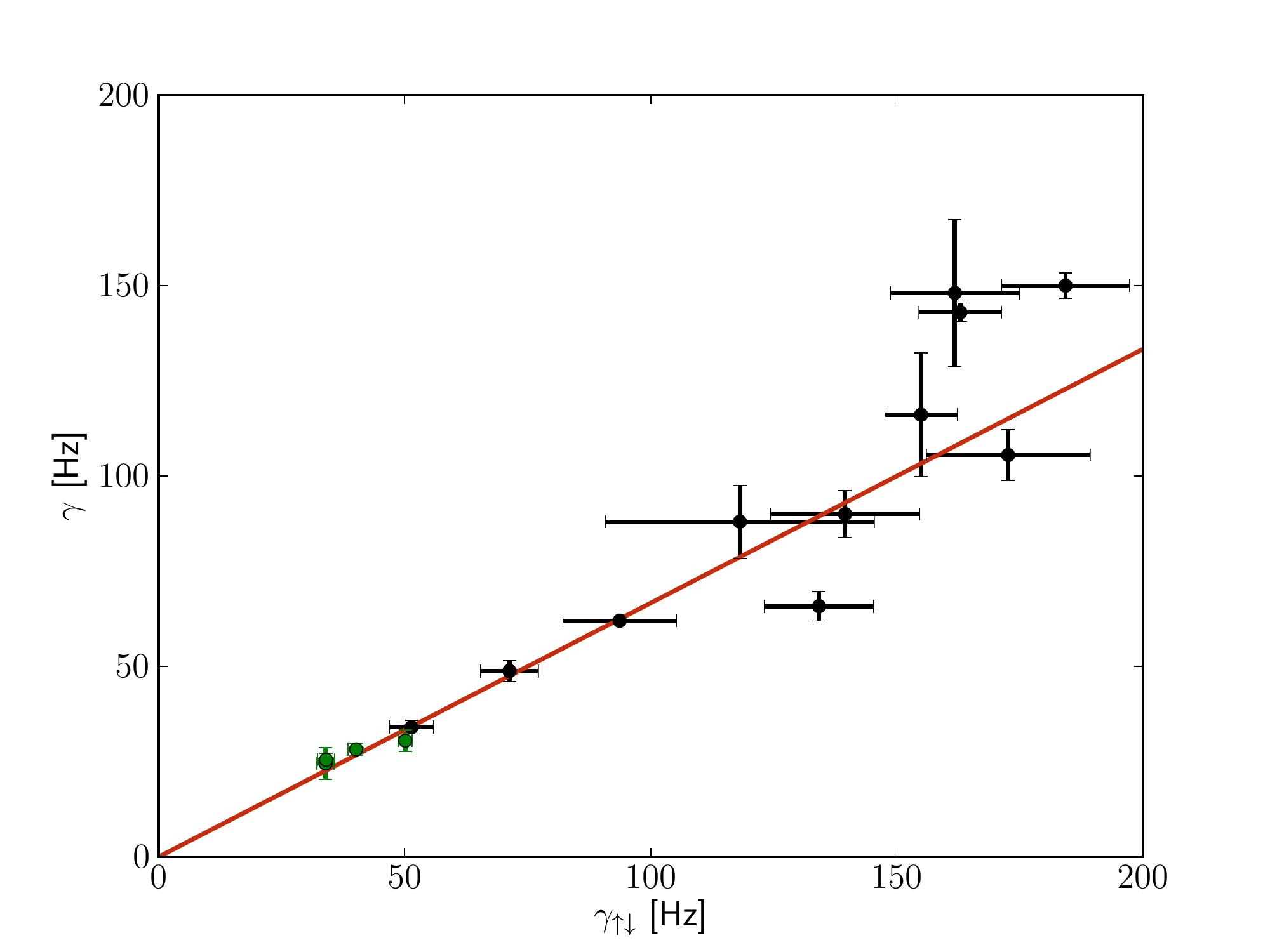}
\end{center}
\caption{{\bf Drag rate versus the interspecies collision rate in the classical regime} -- Drag rate $\gamma$ as a function of the interspecies collision rate $\gamma_{\uparrow\downarrow}$. Here the results for the \drag\ method are indicated with black dots and the results of the \decay\ method are indicated by green dots. The red line represents the prediction of Stringari and Pitaevskii\cite{Stringari} and in this range of $\gamma_{\uparrow\downarrow}$ agrees with our experiment results.  \label{fg:betavsgam}}
\end{figure}

% Drag rate vs. collision rate
As explained earlier, spin drag is caused by collisions between atoms in different spin states. It is therefore to be expected that the drag rate $\gamma$ should be proportional to the interspecies collision rate $\gamma_{\uparrow\downarrow}$. For a non-degenerate gas Stringari and Pitaevskii\cite{Stringari} find $\gamma=\frac{2}{3}\gamma_{\uparrow\downarrow}$. The factor ${2}/{3}<1$, independent of temperature, reflects the fact that for the drag not all collisions contribute equally because only the velocity component in the direction of the force is relevant. In Fig.~\ref{fg:betavsgam} we plot our results for the drag rate as a function of the interspecies collision rate, which is given for the trap by $\gamma_{\uparrow\downarrow} =  \langle n \rangle \sigma_{\uparrow\downarrow} \bar{v}_{\rm rel}$. Here $\langle n \rangle$ is the density of one spin species averaged over the trap, $\sigma_{\uparrow\downarrow}$ is the interspecies collisional cross section and $\bar{v}_{\rm rel}$ the relative velocity between the two colliding atoms averaged over the velocity distribution. As shown in Fig.~\ref{fg:betavsgam}, our results in the non-degenerate regime are in good agreement with the result of Stringari and Pitaevskii\cite{Stringari}.

\begin{figure}
\begin{center}
\includegraphics[width=\textwidth]{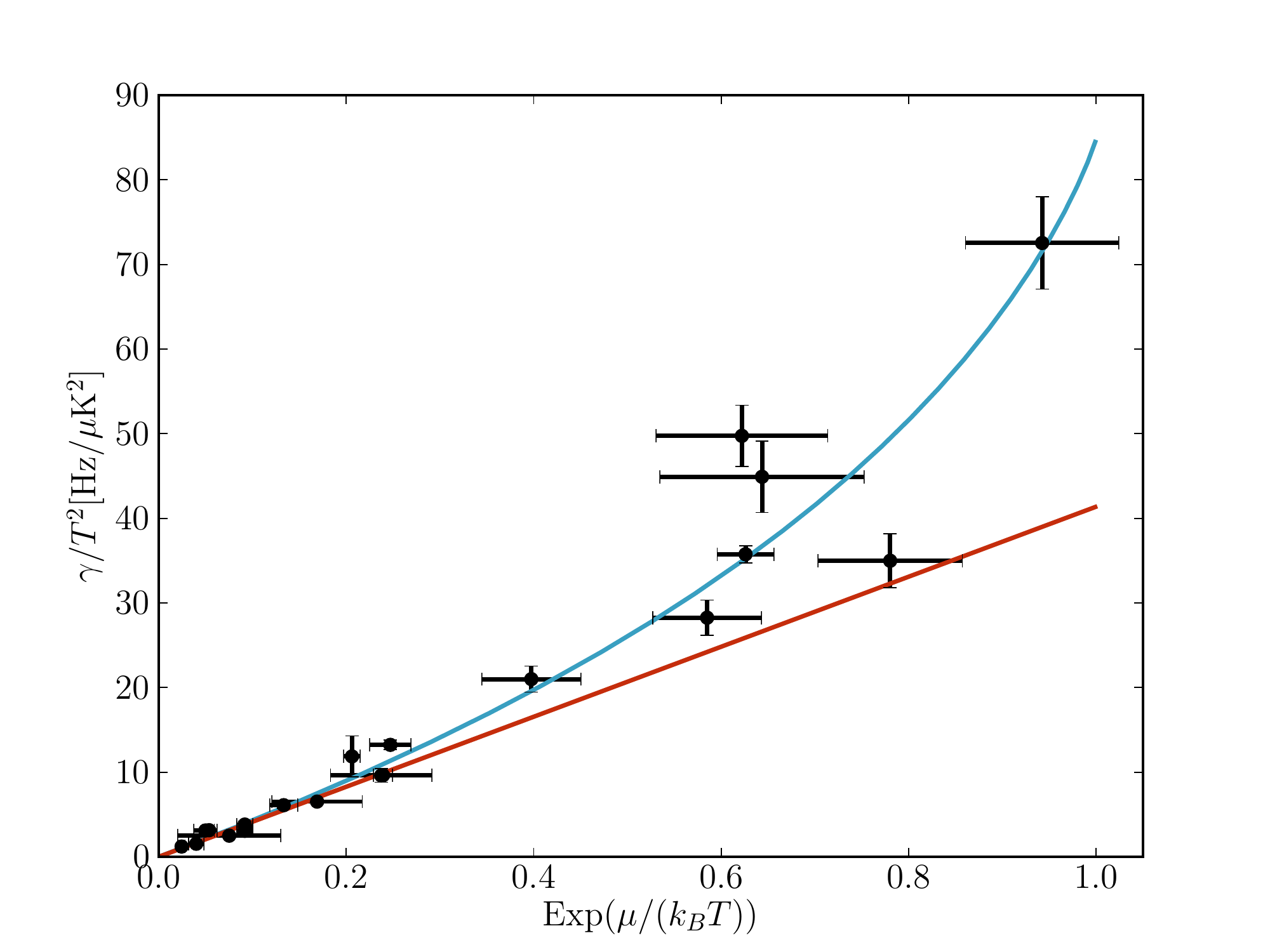}
\end{center}
\caption{{\bf Drag rate versus the fugacity in the degenerate regime} -- Scaled drag rate $\gamma/T^2$ as a function of the fugacity $z=\exp(\mu/k_B)$. The dots represent the data points, whereas the curves show the results for the spin-drag rate using the classical results of Stringari and Pitaevskii\cite{Stringari} (red curve) and the quantum results of van Driel {et al.}\cite{Driel} (blue curve). For small $z$ the gas behaves classically and the data agrees with both models. For large $z$ ($z >$ 0.5) the data deviates from the classical model and is in agreement with the quantum model. The results show that the measurement of the spin-drag rate is a precursor for the phase transition of Bose-Einstein condensation at $z=1$. \label{fg:universal}}
\end{figure}

% Drag rate, degenerate
For larger phase-space densities the gas can no longer be considered as classical and quantum statistics becomes important. For our Bose gas collisions are enhanced at low temperatures due to Bose stimulation. This should be contrasted with fermionic systems, where the drag rate is suppressed at low temperatures  due to Pauli blocking\cite{DeMarco,Sommer}. Collisions are enhanced due to two effects. First, the central densities of the gas in a trap with a fixed number of particles and temperature increases due to Bose enhancement beyond its classical value. Second, the scattering process is enhanced due to Bose stimulation to the final states. In this regime we no longer expect to find a temperature-independent proportionality factor between the drag rate and the interspecies collision rate. Therefore, in Fig.~\ref{fg:universal} the drag rate is shown as a function of the fugacity $z = \exp(\mu/k_B T)$, where $\mu$ is the chemical potential of the gas. For $z \ll 1$ the gas is in the classical limit, whereas for $z=1$ the gas  starts to Bose-Einstein condense. Van Driel {\it et al.}\cite{Driel} showed that the drag rate scaled as $\gamma/T^2$ is a function of the fugacity $z$ only. In Fig.~\ref{fg:universal} we plot   $\gamma/T^2$ as a function of the fugacity $z$, where $z$ and $T$ are determined from the experiment (see Methods). For a classical gas $\gamma/T^2$ depends linearly on $z$ and the classical result is indicated in the figure with a straight line. For small $z$ our measurements agree with this result, but for larger values of $z$ the drag rate increases above the classical rate. Evaluating the expression of van Driel {\it et al.}\cite{Driel} for the drag rate of an inhomogeneous sample, we find the Bose enhancement of the drag rate, as indicated in Fig.~\ref{fg:universal} by the blue line. As can be seen from the figure, our results are in good agreement with theory. As a measure of the accuracy of the theory we use the reduced chi-squared parameter $\chi^2_{\rm red}$, which becomes unity for a large number of observations when the data agrees with the theory. For the classical result we obtain $\chi_{\rm red}^2$ = 6.10, whereas the calculations including Bose enhancement yields $\chi_{\rm red}^2$ = 1.58. We emphasize that here   $\chi_{\rm red}^2$ is used to compare our experimental results to a theory without any adjustable parameters. 

% Hydrodynamic
It is important to realize that the experiments presented here can only be performed when the sample is in the hydrodynamic regime. In this case the number of collisions during one trap oscillation is much larger than one. To quantify this we can define a hydrodynamicity parameter\cite{Gehm} $\phi={\gamma_{\mathrm{col}}}/{\omega_{\mathrm{ax}}}$, with $\gamma_{\rm col}$  the total collision rate including inter- and intraspecies collisions. For the current experiments we have reached $\gamma_{\rm col}$ = 1.4 kHz at an axial trap frequency of $\omega_{\rm ax}/2\pi$ = 3.5 Hz, yielding a hydrodynamicity of $\phi \, >$ 60.  This large hydrodynamicity allows us to quickly reach a steady state. This can for instance be seen in Fig.~\ref{fg:fitcurve}, where a steady state is reached after 5 ms corresponding to only a few collisions. 

% Outlook
In conclusion, we have studied spin drag and determined the spin-drag rate for a gas of ultra-cold bosonic atoms. We find that in the quantum regime spin drag is Bose enhanced and up to almost a factor of two larger than the classical value, which is in good agreement with a recent theoretical prediction\cite{Driel}. This shows that the measurement of spin drag is a strong precursor of Bose-Einstein condensation.  Our results pave the way for studies of the transport properties of degenerate bosons which are very different from fermionic systems, such as electrons in ordinary conductors, and give a complementary picture of transport. Our results are among the first that study transport properties of ultra-cold atoms and in the long term are important for the development of atomtronic devices that are based on transport of ultra-cold atoms\cite{Holland}.

\section{Methods Summary}

Sodium atoms are trapped and cooled in the $|F\!=\!1,m\!=\!-1\rangle\equiv|\uparrow\rangle$ state (pseudo-spin {\em up}) in a magnetic trap, before transferred to a FORT. In the FORT a radio-frequency sweep is applied, such that half of the atoms are transferred to the state $|F\!=\!1,m\!=\!0\rangle\equiv|\downarrow\rangle$ state (pseudo-spin {\em down}). The atoms in the FORT are tightly confined in the radial direction and the sample is hydrodynamic in the axial direction. The force on the atoms is applied using a magnetic field gradient, which only acts on the spin {\em up} atoms. After the experiment the two clouds are spatially separated using a radially inhomogeneous magnetic field and subsequently imaged using absorption imaging. The spin-drag rate is extracted by solving coupled equations for the position of the individual spin species and extracting the relative displacement $\Delta x$ from the solutions where the spin-drag rate is a free parameter. This solution can be compared to the experimental data to extract the spin-drag rate. 

The inter- and intraspecies collision rates are determined using the measured number of atoms in each cloud and the temperature. Combined with the known scattering cross section $\sigma$ and the trap frequencies in the radial and axial direction the inter-, intraspecies and total scattering rate can be determined. %By detecting the number of atoms in spin species {\em up} and {\em down}, the interspecies collision rate can be determined as well. 

\section{Methods}

\subsection{Preparation}
The samples were prepared in a magnetic trap by standard laser and evaporative cooling methods of sodium atoms in the $|F\!=\!1,m\!=\!-1\rangle\equiv|\uparrow\rangle$ state (pseudo-spin {\em up}). There are two major improvements with respect to other experiments concerning the particle numbers: Spin polarizing atoms before the transfer from the magneto-optical trap to the magnetic trap and decompressing the trap axially at the end of the evaporative cooling stage\cite{Stam}. The final trap frequencies are axially $\omega_{\mathrm{ax}}/2\pi$ = 1.5 Hz and radially $\omega_{\mathrm{rad}}/2\pi$ = 116 Hz with up to $8\times10^{8}$ atoms. After evaporatively cooling the atoms to a few $\mu$K the FORT is adiabatically turned on in 400 ms with a trap depth of about 50 $\mu$K, an axial trap frequency of $\omega_{\mathrm{ax}}/2\pi$ = 3.5 Hz and a radial trap frequency of $\omega_{\mathrm{rad}}/2\pi$ = 835 Hz, leading to $4.6\times10^{8}$ atoms in the trap. The lifetime is more than 7 s. The atoms in the $|F\!=\!1,m\!=\!0\rangle\equiv|\downarrow\rangle$ state (pseudo-spin {\em down}) are produced with a radio-frequency sweep from below the bottom of the magnetic trap: 2.6 MHz to  2.71 MHz in 30 to 50 ms. The fraction of atoms in the $|F\!=\!1,m\!=\!+1\rangle$ state is estimated to be less than 10\% and the same for all measurements and does not influence the drag rate.

\subsection{Experiment and Detection}
The force on the spin species {\em up} is produced by a magnetic  field gradient using anti-Helmholtz coils with a constant offset magnetic field such that the zero crossing of the field is not in the area of interest. The detection of different spin species is achieved by turning on the radial magnetic confinement for 5 ms only and then ramping down the FORT in 2 ms for radial cooling. The radial cooling is performed to assure spatial separation of the two spin species. After this time the spin species {\em down} falls freely due to gravitation while the spin species {\em up} is still radially trapped. An additional time of flight of 5--10 ms before imaging is used depending on the density and temperature of the clouds. The imaging is performed by sending a probe beam through the two clouds and detecting the absorption on a CCD camera.

\subsection{Analysis}
The axial position difference is determined by fitting the clouds with a Gaussian distribution multiplied by $1-\mathrm{Erf}(\sigma(x-x_{0}))$, where $\mathrm{Erf}$ is the error function, $x_{0}$ the central position in the Gaussian distribution and $\sigma$ a measure of skewness. From the fits the number of atoms of each species and the temperature are deduced. The latter is deduced from the length of the clouds when the force is applied for 1 ms only. Each measurement is repeated three times for error reduction. Temperature and total number of atoms are also measured separately by detecting a non-perturbed cloud expanding freely in 30 ms time of flight and fitted with a Gaussian distribution. Especially for large clouds this is necessary since the anharmonicity of the trap in the axial direction causes a systematic error in the temperature measurement.

The differential equations describing the system are
\begin{eqnarray}
   \ddot{x_\uparrow} - \gamma \Delta \dot{x} + \frac{1}{\mNa}\frac{\mathrm{d}V(x_\uparrow)}{\mathrm{d}x_\uparrow}&=&\frac{F_\uparrow}{\mNa}\\
   \ddot{x_\downarrow} + \gamma \Delta \dot{x} + \frac{1}{\mNa}\frac{\mathrm{d}V(x_\downarrow)}{\mathrm{d}x_\downarrow}&=&0
\end{eqnarray}
with $x_{\uparrow,\downarrow}$ the axial center-of-mass position of spin species {\em up} and {\em down}, respectively, $\Delta x \equiv x_\uparrow - x_\downarrow$, $\mNa$ the atomic mass, $V(x)$ the axial trapping potential and $F_\uparrow$ the applied force on the spin species {\em up}. Here drag rate $\gamma \equiv \Gamma(\Delta \dot{x})/(f_\uparrow f_\downarrow \, \mNa \, n_{\rm tot} \, \Delta \dot{x}_{\uparrow\downarrow})$ is the spin-drag rate with $\Gamma(v)$ the drag force on the two clouds, $n_{\rm tot}$ the total density of the atoms, and $f_{\uparrow,\downarrow}$ the fraction of atoms with spin {\em up} and {\em down}, respectively. Neglecting the effect of imbalance in $\Gamma(v)$, this expression corrects to leading order for non-equal spin mixtures and neglects higher-order corrections. These equations are numerically integrated using the initial conditions $x_\uparrow(0) = x_\downarrow(0) = 0$ and from the results the difference in position $\Delta x$ as a function of time $t$ is determined. It can be shown by using similar equations for three spin states that the drag rate between two of these species as defined above is not influenced by the presence of a third species, if the three drag rates are the same.

The classical interspecies collision rate is obtained by integrating the interspecies collision rate over the trap and we find\cite{Landaustat1}
\begin{equation}
  \gamma_{\uparrow\downarrow}   = \langle n \rangle \sigma_{\uparrow\downarrow} \bar{v}_{\rm rel} = \frac{N f_\uparrow f_\downarrow \: \mNa \: \sigma_{\uparrow\downarrow} \: \omega_{\mathrm{rad}}^{2} \: \omega_{\mathrm{ax}}}{\pi^2 k_B T}. 
\label{eq:colrate}
\end{equation}
where $\bar{v}_{\rm rel}$ is the relative velocity between the two atoms, $N$ is the total number of atoms and  $\sigma_{\uparrow\downarrow}=4\pi a_{\uparrow\downarrow}{}^{2}$ the interspecies cross section. For $^{23}$Na the scattering length is $a_{\uparrow\downarrow}$ = 2.80 nm\;\cite{Samuelis}.

In the \decay\ method the separation between the clouds is induced using the same magnetic field as in the \drag\ method and the displacement of spin {\em up} is well within the range where the FORT is harmonic. The solution of the equation of motion is
\begin{equation}
  \Delta x(t)= \Delta x(0) \exp\left(-\frac{\beta_d \: t}{2}\right)
\end{equation}
with $\Delta x(0)$ the initial position difference and $\beta_d$ the decay rate. From $\beta_d$ we can deduce the drag rate $\gamma$ using
\begin{equation}
  \gamma = \frac{2 \omega_{\mathrm{ax}}^{2}}{\beta_d}.
\end{equation} 
To determine $\gamma$ theoretically we evaluate the microscopic expression for the spin-drag rate including Bose enhancement given by van Driel {\it et al}\cite{Driel}.

%\bibliography{myrefernces2}

\section{References}
\begin{thebibliography}{10}
\expandafter\ifx\csname url\endcsname\relax
  \def\url#1{\texttt{#1}}\fi
\expandafter\ifx\csname urlprefix\endcsname\relax\def\urlprefix{URL }\fi
\providecommand{\bibinfo}[2]{#2}
\providecommand{\eprint}[2][]{\url{#2}}

\bibitem{Wolf}
\bibinfo{author}{Wolf, S.~A.} \emph{et~al.}
\newblock \bibinfo{title}{Spintronics: A spin-based electronics vision for the
  future}.
\newblock \emph{\bibinfo{journal}{Science}} \textbf{\bibinfo{volume}{294}},
  \bibinfo{pages}{1488--1495} (\bibinfo{year}{2001}).

\bibitem{Mae2002}
\bibinfo{editor}{Maekawa, S.} \& \bibinfo{editor}{Shinjo, T.} (eds.).
\newblock \emph{\bibinfo{title}{Spin Dependent Transport in Magnetic
  Nanostructures, Series: Advances in Condensed Matter Science}}
  (\bibinfo{publisher}{CRC Press}, \bibinfo{year}{2002}).

\bibitem{Zutic}
\bibinfo{author}{\v{Z}uti\'{c}, I.}, \bibinfo{author}{Fabian, J.} \&
  \bibinfo{author}{Das~Sarma, S.}
\newblock \bibinfo{title}{Spintronics: Fundamentals and applications}.
\newblock \emph{\bibinfo{journal}{Rev. Mod. Phys.}}
  \textbf{\bibinfo{volume}{76}}, \bibinfo{pages}{323--410}
  (\bibinfo{year}{2004}).

\bibitem{Aws2007}
\bibinfo{author}{Awschalom, D.~D.} \& \bibinfo{author}{Flatte, M.~E.}
\newblock \bibinfo{title}{Challenges for semiconductor spintronics}.
\newblock \emph{\bibinfo{journal}{Nat. Phys.}} \textbf{\bibinfo{volume}{3}},
  \bibinfo{pages}{153--159} (\bibinfo{year}{2007}).

\bibitem{Ohn2010}
\bibinfo{author}{Ohno, H.}
\newblock \bibinfo{title}{A window on the future of spintronics}.
\newblock \emph{\bibinfo{journal}{Nat. Mater.}} \textbf{\bibinfo{volume}{9}},
  \bibinfo{pages}{952--954} (\bibinfo{year}{2010}).

\bibitem{PhysRevB.68.045307}
\bibinfo{author}{D'Amico, I.} \& \bibinfo{author}{Vignale, G.}
\newblock \bibinfo{title}{Spin coulomb drag in the two-dimensional electron
  liquid}.
\newblock \emph{\bibinfo{journal}{Phys. Rev. B}} \textbf{\bibinfo{volume}{68}},
  \bibinfo{pages}{045307} (\bibinfo{year}{2003}).

\bibitem{Weber}
\bibinfo{author}{Weber, C.~P.} \emph{et~al.}
\newblock \bibinfo{title}{Observation of spin coulomb drag in a two-dimensional
  electron gas}.
\newblock \emph{\bibinfo{journal}{Nature}} \textbf{\bibinfo{volume}{437}},
  \bibinfo{pages}{1330--1333} (\bibinfo{year}{2005}).

\bibitem{DeMarco}
\bibinfo{author}{DeMarco, B.} \& \bibinfo{author}{Jin, D.~S.}
\newblock \bibinfo{title}{Spin excitations in a fermi gas of atoms}.
\newblock \emph{\bibinfo{journal}{Phys. Rev. Lett.}}
  \textbf{\bibinfo{volume}{88}}, \bibinfo{pages}{040405}
  (\bibinfo{year}{2002}).

\bibitem{Sommer}
\bibinfo{author}{Sommer, A.}, \bibinfo{author}{Ku, M.}, \bibinfo{author}{Roati,
  G.} \& \bibinfo{author}{Zwierlein, M.~W.}
\newblock \bibinfo{title}{Universal spin transport in a strongly interacting
  fermi gas}.
\newblock \emph{\bibinfo{journal}{Nature}} \textbf{\bibinfo{volume}{472}},
  \bibinfo{pages}{204} (\bibinfo{year}{2011}).

\bibitem{Driel}
\bibinfo{author}{van Driel, H.~J.}, \bibinfo{author}{Duine, R.~A.} \&
  \bibinfo{author}{Stoof, H. T.~C.}
\newblock \bibinfo{title}{Spin-drag hall effect in a rotating bose mixture}.
\newblock \emph{\bibinfo{journal}{Phys. Rev. Lett.}}
  \textbf{\bibinfo{volume}{105}}, \bibinfo{pages}{155301}
  (\bibinfo{year}{2010}).

\bibitem{Holland}
\bibinfo{author}{Seaman, B.~T.}, \bibinfo{author}{Kr\"amer, M.},
  \bibinfo{author}{Anderson, D.~Z.} \& \bibinfo{author}{Holland, M.~J.}
\newblock \bibinfo{title}{Atomtronics: Ultracold-atom analogs of electronic
  devices}.
\newblock \emph{\bibinfo{journal}{Phys. Rev. A}} \textbf{\bibinfo{volume}{75}},
  \bibinfo{pages}{023615} (\bibinfo{year}{2007}).

\bibitem{Bednorz}
\bibinfo{author}{{Bednorz}, J.~G.} \& \bibinfo{author}{{M{\"u}ller}, K.~A.}
\newblock \bibinfo{title}{{Possible high T$_{c}$ superconductivity in the
  Ba-La-Cu-O system}}.
\newblock \emph{\bibinfo{journal}{Zeitschrift fur Physik B Condensed Matter}}
  \textbf{\bibinfo{volume}{64}}, \bibinfo{pages}{189--193}
  (\bibinfo{year}{1986}).

\bibitem{Klitzing}
\bibinfo{author}{Klitzing, K.~v.}, \bibinfo{author}{Dorda, G.} \&
  \bibinfo{author}{Pepper, M.}
\newblock \bibinfo{title}{New method for high-accuracy determination of the
  fine-structure constant based on quantized hall resistance}.
\newblock \emph{\bibinfo{journal}{Phys. Rev. Lett.}}
  \textbf{\bibinfo{volume}{45}}, \bibinfo{pages}{494--497}
  (\bibinfo{year}{1980}).

\bibitem{Tsui}
\bibinfo{author}{Tsui, D.~C.}, \bibinfo{author}{Stormer, H.~L.} \&
  \bibinfo{author}{Gossard, A.~C.}
\newblock \bibinfo{title}{Two-dimensional magnetotransport in the extreme
  quantum limit}.
\newblock \emph{\bibinfo{journal}{Phys. Rev. Lett.}}
  \textbf{\bibinfo{volume}{48}}, \bibinfo{pages}{1559--1562}
  (\bibinfo{year}{1982}).

\bibitem{Stenger}
\bibinfo{author}{Stenger, J.} \emph{et~al.}
\newblock \bibinfo{title}{Spin domains in ground-state bose-einstein
  condensates}.
\newblock \emph{\bibinfo{journal}{Nature}} \textbf{\bibinfo{volume}{396}},
  \bibinfo{pages}{345--348} (\bibinfo{year}{1998}).

\bibitem{Duine}
\bibinfo{author}{Duine, R.~A.} \& \bibinfo{author}{Stoof, H.}
\newblock \bibinfo{title}{Spin drag in noncondensed bose gases}.
\newblock \emph{\bibinfo{journal}{Phys. Rev. Lett.}}
  \textbf{\bibinfo{volume}{103}}, \bibinfo{pages}{170401}
  (\bibinfo{year}{2009}).

\bibitem{Stringari}
\bibinfo{author}{Pitaevskii, L.} \& \bibinfo{author}{Stringari, S.}
\newblock \emph{\bibinfo{title}{Bose-Einstein Condensation}}
  (\bibinfo{publisher}{Oxford Univ. Press}, \bibinfo{address}{Oxford},
  \bibinfo{year}{2003}).

\bibitem{Gehm}
\bibinfo{author}{Gehm, M.~E.}, \bibinfo{author}{Hemmer, S.~L.},
  \bibinfo{author}{O'Hara, K.~M.} \& \bibinfo{author}{Thomas, J.~E.}
\newblock \bibinfo{title}{Unitarity-limited elastic collision rate in a
  harmonically trapped fermi gas}.
\newblock \emph{\bibinfo{journal}{Phys. Rev. A}} \textbf{\bibinfo{volume}{68}},
  \bibinfo{pages}{011603} (\bibinfo{year}{2003}).

\bibitem{Stam}
\bibinfo{author}{van~der Stam, K. M.~R.}, \bibinfo{author}{Ooijen, E.~D.},
  \bibinfo{author}{Meppelink, R.}, \bibinfo{author}{Vogels, J.~M.} \&
  \bibinfo{author}{van~der Straten, P.}
\newblock \bibinfo{title}{Large atom number bose-einstein condensates of
  sodium}.
\newblock \emph{\bibinfo{journal}{Rev. Sci. Instrum.}}
  \textbf{\bibinfo{volume}{78}}, \bibinfo{pages}{013102}
  (\bibinfo{year}{2007}).

\bibitem{Landaustat1}
\bibinfo{author}{Landau, D., L.} \& \bibinfo{author}{Lifshitz, E.}
\newblock \emph{\bibinfo{title}{Course of Theoretical Physics, Volume 5:
  Statistical physics Part 1}} (\bibinfo{publisher}{Butterworth-Heinemann},
  \bibinfo{address}{Oxford}, \bibinfo{year}{1999}).

\bibitem{Samuelis}
\bibinfo{author}{Samuelis, C.} \emph{et~al.}
\newblock \bibinfo{title}{Cold atomic collisions studied by molecular
  spectroscopy}.
\newblock \emph{\bibinfo{journal}{Phys. Rev. A}} \textbf{\bibinfo{volume}{63}},
  \bibinfo{pages}{012710} (\bibinfo{year}{2000}).

\end{thebibliography}

\begin{addendum}
\item[Acknowledgements] We would like to thank D. van Oosten for stimulating discussions during the experiments and H. van Driel and R. Kittinaradorn for help with the numerical calculations. 
  \item[Author contribution] All authors contributed extensively to the work presented in this paper.
  
% S.B. Koller and A. Groot built the experiment and with P. Bons took the measurements. 
%Analysis was done by S.B. Koller and P. van der Straten. 
%Calculations are performed by P. van der Straten, R.A. Duine and H.T.C. Stoof 
%The manuscript is written by S.B. Koller P. van der Straten.  
%All authors discussed the measurements and manuscript. 

\end{addendum}

\end{document}